\documentclass[prc,showpacs,floatfix]{revtex4}
\usepackage{bm}
\usepackage{dcolumn}
\usepackage{graphicx}
\usepackage{amsmath}
\newcommand{\beq}{\begin{equation}}
\newcommand{\eeq}{\end{equation}}
\newcommand{\beqa}{\begin{eqnarray}}
\newcommand{\eeqa}{\end{eqnarray}}
\begin{document}
\title{
\hfill{\small {\bf MKPH-T-12-07}}\\
{\bf Partial wave expansion for photoproduction of two pseudoscalars on a nucleon}}
\author{
A.~Fix$^{1}$ and H.~Arenh\"ovel$^{2}$}
\affiliation{
$^1$Laboratory of Mathematical Physics, Tomsk Polytechnic University, 634034 Tomsk, Russia\\
$^2$Institut f\"ur Kernphysik, Johannes Gutenberg-Universit\"at Mainz,
D-55099 Mainz, Germany}
\date{\today}
\begin{abstract}
The amplitudes for photoproduction of two pseudoscalars on a nucleon are expanded in the
overall c.m.~frame in a model independent way with respect to the contribution of the
final state partial wave of total angular momentum $J$ and its projection on the normal
to the plane spanned by the momenta of the final particles. The expansion coefficients
which are analogues to the multipole amplitudes for single meson photoproduction contain
the complete information about the reaction dynamics. Results of an explicit evaluation
are presented for the moments $W_{jm}$ of the inclusive angular distribution of an
incident photon beam with respect to the c.m.\ coordinate system defined by the final
particles taking photoproduction of $\pi^0\pi^0$ and $\pi^0\eta$ as an example.
\end{abstract}

 \maketitle


\section{Introduction}
The study of multiple meson production is essential for understanding the properties of
baryonic resonances, especially of those having sizeable inelasticities and for which
only a weak evidence from elastic $\pi N$ scattering exists. According to the quark model
calculation of~\cite{Roberts1}, at least below 2 GeV some of these resonances must be
strongly coupled to $\pi\pi N$ and $\pi\eta N$ channels. Therefore,
present experiments on $\pi\pi$
and $\pi\eta$ photoproduction have become a center of attention in
programs discussed at various research centers, and a number of new
accurate data have already been
reported~\cite{Horn,Weinheimer,Tohoku,Ajaka,Kashev,Gutz_Is,GutzBeam,KashevAphi}.

Improvements in the quality of the data have made it possible to perform rather detailed
theoretical analyses of photoproduction of two pseudoscalars. The $\pi\pi$ as well as
$\pi\eta$ models have already been the object of several
studies~\cite{Oset,Laget,Ochi,Mokeev,FA,Dor,Sara,FKLO,Doring}. Mainly they cover the second
and third resonance regions describing with varying degrees of success the
existing data and predicting the results of new measurements. A typical analysis is based
on an isobar model approach. Its key assumption is that the amplitude is a coherent sum
of background and resonances usually parametrized in terms of effective
Lagrangeans. As a rule, the resonance part contains $s$-channel resonances decaying into
$\pi\pi N$ or $\pi\eta N$ via intermediate formation of meson-nucleon and meson-meson
isobars. As adjustable parameters one usually takes the masses and partial decay widths
of the resonances as well as their electromagnetic coupling constants.

Within this method, angular momentum decomposition of the amplitude is ruled by partial
wave transitions of the resonance states to quasi-two-body states, like $\pi\Delta$ or
$\eta\Delta$. Clearly, such an approach can not be viewed as a general partial wave
analysis, since it crucially depends on the assumptions about the
production mechanism. Therefore, it results in various
uncertainties, primarily in the non-uniqueness of existing
solutions, since the same observables may equally well be described with different sets
of parameters. Consequently, in spite of a general qualitative agreement among the
models, significant quantitative discrepancies still remain. Additional limitations may
arise from inadequacies of the isobar model description, such as violation of unitarity,
nonrelativistic dynamics {\it etc.}, whose impact on the description of the processes
under discussion remains unknown. For instance, unitarity conditions may be important in
the region where many production channels are open.

It is worth to note that one of the main reasons for the lack of a rigorous partial wave
analysis for $\pi\pi$ and $\pi\eta$ photoproduction is that there is no general recipe to
deal with reactions involving three particles in the final state. In contrast to single
meson photoproduction one faces here the technical problems associated with three-body
kinematics, where the particle energies and angles are distributed continuously. As a
consequence, a conventional partial wave decomposition of the final state does not provide
a multipole representation for practical applications, primarily since there exists a variety
of ways to successively couple angular momenta of the participating particles to a total angular
momentum.

In this paper we present an alternative method by using a partial wave expansion for the
photon induced production of two pseudoscalars on a nucleon, which should be of minimal
model dependence. It is based on the correct determination of the partial wave amplitudes
for these reactions with no built-in prejudices concerning the production mechanism.
Similar method have been used to analyse pion production in $\pi N$ collisions, see,
for example Refs.~\cite{Arnold,Morgan}.

The paper is organized as follows. In the next section we introduce
the partial wave expansion and construct the transition
amplitude for photoproduction of two pseudoscalar mesons. In Sect.~\ref{2pi0} we use the
so far developed formalism to discuss some gross features of $\pi^0\pi^0$ and $\pi^0\eta$
photoproduction. Finally, some general conclusions are drawn in Sect.~\ref{conclusion}.

\section{The Formalism}\label{formal}
In this section, we collect the formulas used in the present
analysis. As a starting point the formal results
of Ref.~\cite{FiA11} are used. There the formal expressions for the helicity
amplitudes as well as for the cross section and the recoil polarization were derived,
including various polarization asymmetries with respect to polarized photons and
nucleons.

\subsection{The T matrix}
We consider here the photoproduction of two pseudoscalar mesons, denoted $m_1$ and
$m_2$ with masses $M_1$ and $M_2$, respectively. Firstly we determine the $T$-matrix
elements of the electromagnetic $m_1m_2$ production current $\vec
J_{\gamma m_1m_2}$
between the initial nucleon and the final $m_1 m_2 N$ state. The four-momenta of incoming
photon, outgoing mesons, initial and final nucleons are denoted by
$(\omega_\gamma,\vec{k}\,)$, $(\omega_1,\vec{q}_1\,)$, $(\omega_2,\vec{q}_2\,)$,
$(E_i,\vec{p}_i\,)$, and $(E,\vec{p}\,)$, respectively. The helicities of photon and
initial and final nucleons are denoted by $\lambda$, $\mu$, and $\nu$,
respectively. In a general frame the transition matrix element is given by
\begin{equation}
T_{\nu\lambda\mu}= -^{(-)}\langle \vec p,\, \vec q,\,\nu\,|\
\vec\varepsilon_\lambda\cdot\vec J_{\gamma m_1m_2}(0)|\,\vec p_i,\mu\rangle\,,
\end{equation}
where for the description of the final state we choose the final nucleon momentum $\vec
p=(p,\theta_p,\phi_p)$ and the relative momentum of the two mesons $\vec
q=\frac12(\vec{q}_1-\vec{q}_2\,) =(q,\theta_q,\phi_q)$. For the following formal
considerations the knowledge of the specific form of the current $\vec
J_{\gamma m_1m_2}$
is not needed.

After separation of the overall c.m.-motion the general form of the
$T$-matrix is given by
\begin{eqnarray}\label{1}
T_{\nu\lambda\mu}&=& - ^{(-)}\langle
\vec{p},\,\vec{q},\,\nu\,|J_{\gamma m_1m_2,\,\lambda}(\vec k\,)|\mu\rangle\,.
\end{eqnarray}
It is convenient to introduce a partial wave decomposition of the
outgoing final state according to
\begin{eqnarray}
^{(-)}\langle \vec q,\, \vec p,\, \nu|=\frac{1}{4\pi} \sum_{l_p j_p
  m_p l_q m_q J M}&&\widehat l_p\, \widehat l_q\,
(l_p0 \frac{1}{2} \nu|j_p \nu)\,(j_p m_p l_q m_q|J M)\, D^{j_p}_{\nu
m_p}(\phi_p,-\theta_p,-\phi_p)\nonumber\\&&\times\,D^{l_q}_{0
m_q}(\phi_q,-\theta_q,-\phi_q)\, ^{(-)}\langle q p;\, ((l_p  \frac{1}{2})j_p l_q)J M|\,,
\end{eqnarray}
where the ``hat'' symbol means, for example, $\widehat
l_q=\sqrt{2l_q+1}$. Furthermore, $l_q$ and $m_q$ denote total angular
momentum and projection, respectively, of the two mesons,  $l_p$,
$j_p$, and $m_p$ orbital and total nucleon angular momentum and its
projection, respectively, and $J$ and $M$ the total angular momentum of the
partial wave and its projection. All projections refer to a
quantization axes to be determined later. For the
rotation matrices $D^j_{m'm}$ we follow the convention of Rose~\cite{Ros57}.

The multipole decomposition of the current reads with $\vec k=(k,\theta_\gamma,\phi_\gamma)$
\begin{equation}
J_{\gamma m_1m_2,\lambda}(\vec{k}\,)=-\sqrt{2\pi}\sum\limits_{LM_L}i^L\widehat{L}\,{\cal
O}^{\lambda L}_{M_L}(k)\, D^L_{M_L \lambda }(\phi_\gamma,\theta_\gamma,-\phi_\gamma)\,,
\end{equation}
where ${\cal O}^{\lambda L}_{M_L}$ contains the transverse electric
and magnetic multipoles
\begin{eqnarray}
{\cal O}^{\lambda L}_{M_L}&=& E_{M_L}^L +\lambda M_{M_L}^L\,.
\end{eqnarray}
For the initial nucleon state we have
\begin{equation}
|\frac12 \mu\rangle = (-1)^{\frac12+\mu}\sum\limits_{m=\pm1/2}|\frac12 m\rangle
\,D^{1/2}_{m -\mu }(\phi_\gamma,\theta_\gamma,-\phi_\gamma)\,.
\end{equation}
Using the Wigner-Eckart theorem and the sum rule for rotation matrices
\begin{equation}
\sum\limits_{M_Lm}\left(\begin{array}{ccc} J & L & \frac12 \cr  -M & M_L & m\cr
\end{array}\right)
D^{1/2}_{m\,-\mu}(R) D^{L}_{M_L\,\lambda}(R)= (-1)^{\lambda-\mu-M}
\left(\begin{array}{ccc} J & L & \frac12 \cr  \mu-\lambda & \lambda & -\mu \cr
\end{array}\right)D^J_{ M\,\lambda-\mu}(R)\,,
\end{equation}
one obtains
\begin{eqnarray}\label{4}
T_{\nu \lambda\mu}= \frac{(-1)^{\nu+\lambda}}{2\,\sqrt{2\pi}} \sum_{L l_p j_p m_p l_q m_q
J M }&&(-1)^{l_p+j_p+l_q+J-M}\,i^L\,\widehat L\,\widehat J\,\widehat l_q \,\widehat
l_p\,\widehat j_p\left(\begin{array}{ccc} l_p &
    \frac{1}{2} & j_p \cr 0 & \nu & -\nu\cr
\end{array}\right)\nonumber\\
&&\times \left(\begin{array}{ccc} j_p & l_q & J \cr m_p & m_q & -M\cr
\end{array}\right)
\left(\begin{array}{ccc} J & L & \frac{1}{2} \cr \mu-\lambda & \lambda & -\mu\cr
\end{array}\right)
\langle p \,q; \big((l_p\frac12)j_p l_q\big)J||{\cal O}^{\lambda L}||\frac{1}{2}\rangle\nonumber\\
&& \times \,D^{j_p}_{\nu\, m_p}(\phi_p,-\theta_p,-\phi_p)\,D^{l_q}_{0\,
m_q}(\phi_q,-\theta_q,-\phi_q)\,
D^{J}_{M\,\lambda-\mu}(\phi_\gamma,\theta_\gamma,-\phi_\gamma)\,.
\end{eqnarray}
Parity conservation results in the following symmetry relation
\begin{equation}\label{ParityT}
T_{-\nu-\lambda-\mu}(\Omega_q,\Omega_p,\Omega_\gamma)=
(-1)^{\lambda-\mu-\nu}T_{\nu \lambda\mu}(\bar\Omega_q,\bar\Omega_p,\bar\Omega_\gamma)\,,
\end{equation}
where for $\Omega=(\theta,\phi)$ we have introduced the notation $\bar\Omega=(\theta,-\phi)$\,.

\begin{figure}
\begin{center}
\includegraphics[scale=.8]{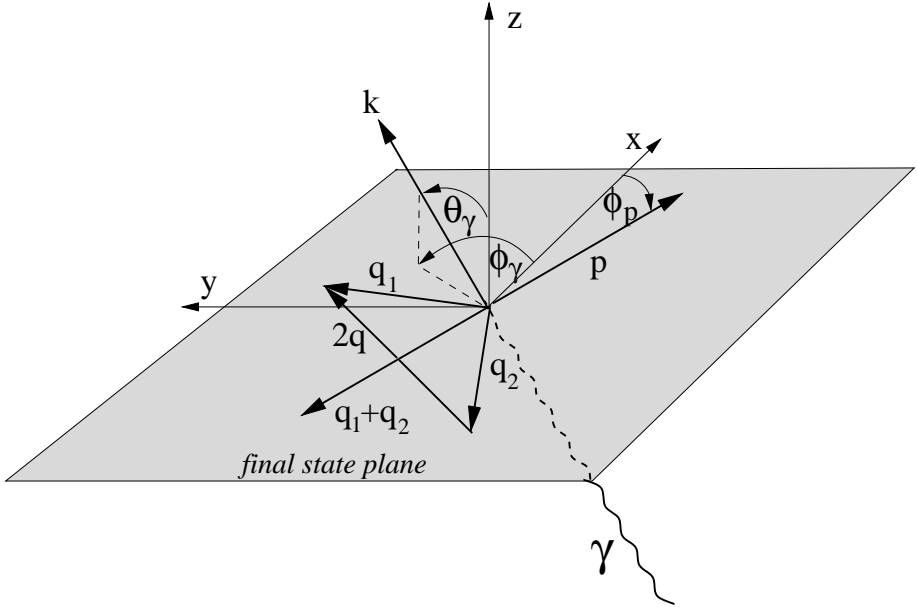}
\caption{Definition of the coordinate system in the c.m.\ system. } \label{fig1}
\end{center}
\end{figure}

Now we turn to the choice of our coordinate system in the overall center-of-momentum frame.
We use the so-called "rigid body" system $K_{fs}$, associated with the
final state plane spanned by the final three particles, in which the
$z$-axis is taken to be the normal to this plane and parallel to $\vec p\times\vec q_1$. Thus
the  $x$- and $y$-axes are in the final scattering plane (see
Fig.~\ref{fig1}).

At a given three-particle invariant energy $W$, the relative
orientation of the final particles within
the final state plane is
characterized by three independent variables for which we take the
angle $\phi_p$ of the final nucleon momentum and the energies of the
two mesons, $\omega_1$ and $\omega_2$ (see Fig.~\ref{fig1}).
After straightforward algebra one obtains for the final nucleon momentum $p$
\begin{equation}\label{10}
p=|\vec{p}\,|=\sqrt{(W-\omega_1-\omega_2)^2-M_N^2}\,,
\end{equation}
and for the relative momentum $q$ of the two mesons
\beq \label{11}
q^2=\frac{1}{2}(\omega_1^2+\omega_2^2-M_1^2-M_2^2)-\frac{p^2}{4}\,.
\eeq

The orientation of the chosen coordinate system with respect to the beam axes may be
specified by $\Omega_\gamma=(\phi_\gamma,\theta_\gamma)$, the spherical angles of the
photon momentum $\vec{k}$ with respect to $K_{fs}$. One readily notes
that in this coordinate system one has $\theta_p=\theta_q=\pi/2$ and therefore
\begin{eqnarray}
D^{j_p}_{\nu
  m_p}(\phi_p,-\theta_p,-\phi_p)&=&(-1)^{\nu-m_p}d^{j_p}_{\nu  m_p}(\pi/2)
\,e^{-i(\nu-m_p)\phi_p}\,,\\
D^{l_q}_{0 m_q}(\phi_q,-\theta_q,-\phi_q)&=&(-1)^{m_q} d^{l_q}_{0 m_q}(\pi/2) \,e^{im_q\phi_q}\,.
\end{eqnarray}
As will be shown soon, instead of $\phi_q$ only
$\phi_{qp}=\phi_q-\phi_p$ is needed.
It is related to $\omega_1$ and $\omega_2$ by
\beq
\cos\phi_{qp}= \frac{1}{2qp}(\omega_2^2-\omega_1^2-M_2^2+M_1^2)\,,
\eeq
with $p$ and $q$ from Eqs.~(\ref{10}) and (\ref{11}), respectively.
Thus we will take as independent variables besides the photon
angles $\Omega_\gamma=(\theta_\gamma,\phi_\gamma)$  and $\phi_p$ the
energies of the two mesons $\omega_1$ and $\omega_2$ instead of $p$
and $\phi_{qp}$ and obtain the following representation of the $T$-matrix
element making the angular dependence explicit
\begin{equation}\label{14}
T_{\nu \lambda\mu}(\phi_p,\omega_1,\omega_2, \Omega_\gamma)=
e^{i(\lambda-\mu)\phi_\gamma}\,e^{-i\nu\phi_p}\sum\limits_{JM}
t^{JM}_{\nu\lambda\mu}(\omega_1,\omega_2) \,e^{-iM\phi_{\gamma p}}
d^J_{M\,\lambda-\mu}(\theta_\gamma)\,,
\end{equation}
with the contribution of the final partial wave
\begin{eqnarray}
t^{JM}_{\nu \lambda\mu}(\omega_1,\omega_2)&=&
t^{JM}_{\nu \lambda\mu} (\phi_{qp})\nonumber\\
&=&\sum_{l_p j_p m_p L}
\left(\begin{array}{ccc} l_p &
    \frac{1}{2} & j_p \cr 0 & \nu & -\nu\cr
\end{array}\right)
\left(\begin{array}{ccc} J & L & \frac{1}{2} \cr \mu-\lambda & \lambda & -\mu\cr
\end{array}\right)
d^{j_p}_{\nu\, m_p}(\pi/2) \,
e^{i(M-m_p)\phi_{qp}}\,{\cal O}^{\lambda LJ}_{M}(l_p j_p m_p)\,,\label{multipole}
\end{eqnarray}
which shows the explicit dependence on $\phi_{qp}$.
Furthermore, we have introduced for convenience the notation
\beqa
{\cal O}^{\lambda LJ}_{M}(l_p j_p m_p) &=&
\frac{(-1)^{1+J}\widehat J}{2\sqrt{2\pi}} \sum_{l_q m_q} i^L (-1)^{l_p+j_p+l_q}
\,\widehat l_p\,\widehat j_p\,\widehat l_q\,\widehat L \,d^{l_q}_{0\,
m_q}(\pi/2)\nonumber\\&&\times
\left(
\begin{matrix}
j_p& l_q&J \cr m_p&m_q&-M \cr
\end{matrix} \right)
\langle p \,q; \big((l_p\frac12)j_p l_q\big)J||{\cal O}^{\lambda
  L}||\frac{1}{2}\rangle\,.
\eeqa
The following symmetry properties hold for the ${\cal O}^{\lambda
  LJ}_{M}(l_p j_p m_p) $
\beqa
{\cal O}^{-\lambda LJ}_{M}(l_p j_p m_p)&=&(-)^{L+l_p +M-m_p}
{\cal O}^{\lambda LJ}_{M}(l_p j_p m_p)\label{symOa}\,,\\
{\cal O}^{\lambda LJ}_{-M}(l_p j_p -m_p)&=&(-)^{j_p+J}
{\cal O}^{\lambda LJ}_{M}(l_p j_p m_p) \label{symOb}\,,
\eeqa
where the first one is a consequence of parity conservation.

The symmetry relation of Eq.~(\ref{ParityT}) leads to the following symmetry property of
the amplitudes $t_{\nu\lambda\mu}^{JM}$
\begin{equation}
t_{-\nu -\lambda-\mu}^{JM}(\phi_{qp})
=(-1)^{\nu+M}t_{\nu\lambda\mu}^{J-M}(-\phi_{qp})\,.\label{sym-t}
\end{equation}
This means that for each $J$ the number of independent amplitudes is $4(2J+1)$.

The complex functions $t_{\nu\lambda\mu}^{JM}$, depending on the meson energies
$\omega_1$ and $\omega_2$ only, provide a complete description of the
process in a manner analogous
to the description of a single meson photoproduction in terms of multipoles. It is worth
to point out, that in contrast to the binary reactions the partial amplitudes are
functions of the c.m.\ energies of the final particles and, therefore, are to be
determined for every point of the Dalitz plot.

\subsection{The differential cross section}\label{diff-cross}

For the unpolarized differential cross section one obtains with the
$T$-matrix of Eq.~(\ref{14})
\begin{eqnarray}
\frac{d^4\sigma_0}{
  d\omega_1d\omega_2d\cos\theta_\gamma d\phi_{\gamma p}}&=&
c(W)\frac{1}{4}\sum_{\nu\lambda\mu}
\left|T_{\nu\lambda \mu}\right|^2\nonumber\\
&=&
\sum_{jm} S_{jm}(\omega_1,\omega_2) Y_{jm}(\theta_\gamma ,\phi_{\gamma p}) \label{dsig0}
\end{eqnarray}
where we have defined
\begin{eqnarray}
 S_{jm}\,c(W)(\omega_1,\omega_2)
&=&\frac{\sqrt{\pi}}{2}\,c(W)\,\widehat
j\sum_{ J'M'JM } (-1)^{-M'} 
\left(
\begin{matrix}
J'& J&j \cr M'&-M&-m \cr
\end{matrix} \right) \nonumber\\ &&\times
\sum_{\nu\lambda\mu} (-1)^{\lambda-\mu}\left(
\begin{matrix}
J'& J&j \cr \lambda-\mu&\mu-\lambda& 0 \cr
\end{matrix} \right)
 t^{J'M'}_{\nu\lambda\mu}(\omega_1,\omega_2)^*\,t^{JM}_{\nu
   \lambda\mu}(\omega_1,\omega_2)
\,,\label{sjm}
\end{eqnarray}
with 
\beq
c(W)=\frac{M_N^2}{4(2\pi)^4(W^2-M_N^2)}
\eeq 
as a kinematical factor.  One should
note that the differential cross section depends on the relative angle
$\phi_{\gamma p}$ only besides on $\omega_1$, $\omega_2$, and
$\theta_\gamma$ as is immediately evident in the absence of
polarization effects.

In terms of the electromagnetic multipole contributions one finds
\beqa
 S_{jm}(\omega_1,\omega_2)
&=&\frac{\sqrt{\pi}}{2}\,c(W)\,\widehat j
\sum_{J_p M_p}\widehat J_p^{\,2}\,d^{J_p}_{0M_p}(\pi/2) \,e^{i(m+M_p)\phi_{q p}}\nonumber\\&&
\times\sum_{l_p' j_p' m_p' l_p j_p m_p}(-1)^{j_p'-j_p -m_p}
\left(
\begin{matrix}
l_p'& l_p&J_p \cr 0&0&0 \cr
\end{matrix} \right)
\left(
\begin{matrix}
j_p'& j_p&J_p \cr m_p'&-m_p&-M_p \cr
\end{matrix} \right)
\left\{
\begin{matrix}
l_p'& l_p&J_p \cr j_p&j_p'&\frac12 \cr
\end{matrix} \right\}\nonumber\\&&
\times\sum_{ J'M'JML'L} (-1)^{J'+J+M'+L'+L'}
\left(
\begin{matrix}
J'& J&j \cr M'&-M&-m \cr
\end{matrix} \right)
\left\{
\begin{matrix}
J'& J&j \cr L&L'&\frac12 \cr
\end{matrix} \right\}\nonumber\\&&
\times\sum_\lambda(-)^\lambda
\left(
\begin{matrix}
L& L'&j \cr \lambda&- \lambda&0 \cr
\end{matrix} \right)
\,{\cal O}^{\lambda L'J'}_{M'}(l_p' j_p' m_p') ^*
\,{\cal O}^{\lambda LJ}_{M}(l_p j_p m_p)\,.
\eeqa

If with respect to the fixed final state plane only the direction of
the final nucleon is detected, one obtains a semi-inclusive
differential cross section by integrating the expression in
Eq.~(\ref{dsig0}) over $\omega_1$ and $\omega_2$ (setting
without loss of generality $\phi_p=0$, which means that $\phi_\gamma$
is measured relative to the direction of the nucleon momentum)
\beqa
d\sigma_2/d\Omega_\gamma &=&\int d\omega_1d\omega_2
\frac{d^4\sigma_0}{  d\omega_1d\omega_2d\Omega_\gamma}\nonumber\\
&=&\sum_{jm} \widetilde S_{jm}Y_{jm}(\Omega_\gamma) \label{diff-semi}
\eeqa
as an expansion in terms of spherical harmonics in $\Omega_\gamma$
with
\beqa
\widetilde S_{jm}&=&\int d\omega_1d\omega_2 S_{jm}(\omega_1,\omega_2)
\nonumber\\
&=&\frac{\sqrt{\pi}}{2}\,c(W)\,\widehat
j\sum_{ J'M'JM } (-1)^{-M'} 
\left(
\begin{matrix}
J'& J&j \cr M'&-M&-m \cr
\end{matrix} \right) \nonumber\\ &&\times
\sum_{\nu\lambda\mu} (-1)^{\lambda-\mu}\left(
\begin{matrix}
J'& J&j \cr \lambda-\mu&\mu-\lambda& 0 \cr
\end{matrix} \right)
 \int d\omega_1d\omega_2 
t^{J'M'}_{\nu\lambda\mu}(\omega_1,\omega_2)^*\,t^{JM}_{\nu
   \lambda\mu}(\omega_1,\omega_2)
\,,\label{tilde-sjm}
\eeqa
or in terms of the multipoles
\beqa
\widetilde S_{jm}&=&
\frac{\sqrt{\pi}}{2}\,c(W)\,\widehat j
\sum_{J_p M_p}\widehat J_p^2\,d^{J_p}_{0M_p}(\pi/2)
\sum_{l_p' j_p' m_p' l_p j_p m_p}(-1)^{j_p'-j_p -m_p}
\left(
\begin{matrix}
l_p'& l_p&J_p \cr 0&0&0 \cr
\end{matrix} \right)
\left(
\begin{matrix}
j_p'& j_p&J_p \cr m_p'&-m_p&-M_p \cr
\end{matrix} \right)
\left\{
\begin{matrix}
l_p'& l_p&J_p \cr j_p&j_p'&\frac12 \cr
\end{matrix} \right\}\nonumber\\&&
\times\sum_{ J'M'JML'L} (-1)^{J'+J+M'+L'+L'}
\left(
\begin{matrix}
J'& J&j \cr M'&-M&-m \cr
\end{matrix} \right)
\left\{
\begin{matrix}
J'& J&j \cr L&L'&\frac12 \cr
\end{matrix} \right\}\nonumber\\&&
\times\sum_\lambda(-)^\lambda
\left(
\begin{matrix}
L& L'&j \cr \lambda&- \lambda&0 \cr
\end{matrix} \right)
\int d\omega_1d\omega_2
\,e^{i(m+M_p)\phi_{q p}}\,{\cal O}^{\lambda L'J'}_{M'}(l_p' j_p' m_p') ^*
\,{\cal O}^{\lambda LJ}_{M}(l_p j_p m_p)\,.
\eeqa

Since $d^2\sigma_0/d\Omega_\gamma$ is a real quantity, one has the
property
\beq
\widetilde S_{jm}^*=(-)^m \widetilde S_{j-m}\,.\label{sjm*}
\eeq
Furthermore, the cross section should be invariant under the
simultaneous inversion of $\vec k$ and $\vec p$, i.e.\
under the transformation $\theta_{\gamma p}\to \pi-\theta_{\gamma p}$. Thus one
finds as additional symmetry property
\beq
\widetilde S_{jm}=(-)^{j+m}\widetilde S_{jm}\,,
\eeq
from which the selection rule $\widetilde S_{jm}=0$ for
$j+m=$\,odd follows. This property can also be shown straightforwardly
using Eq.~(\ref{tilde-sjm}) with the help of Eq.~(\ref{symOb}). For
identical mesons, one finds from Eq.~(\ref{sym-t}) an additional symmetry, namely
\beq
\widetilde S_{j-m}=(-)^{m}\widetilde S_{jm}\,,\label{sjm-ident}
\eeq
which leads in conjunction with Eq.~(\ref{sjm*}) to $\Im m\,\widetilde S_{jm}=0$.

It is more convenient to use instead of the differential
cross section the corresponding normalized quantity
\begin{equation}\label{Wij}
W(\Omega_\gamma)\equiv\frac{1}{\sigma_0}\frac{d^2\sigma_0}{d\Omega_\gamma}=\frac{1}{4\pi}+
\sum_{jm,j\geq 1, j+m=\mathrm{even}}\frac{\widehat{j}}{\sqrt{4\pi}}\,W_{jm}Y_{jm}(\Omega_\gamma)\,,
\end{equation}
where the total cross section $\sigma_0$ is given by
\beqa
\sigma_0&=&2\sqrt{\pi}\widetilde S_{00}
\nonumber\\
&=&\pi\,c(W)\int d\omega_1d\omega_2  \sum_{\nu\lambda\mu J
M}\frac{1}{2J+1}\,|t^{JM}_{\nu \lambda\mu}(\omega_1,\omega_2) |^2\,,
\eeqa
and the expansion coefficients by
\begin{eqnarray}\label{Wjm}
W_{jm}&=&\frac{2\sqrt{\pi}}{\sigma_0 \,\widehat j}\widetilde S_{jm}
\nonumber\\
&=&\frac{\pi}{\sigma_0}\,c(W)\int d\omega_1d\omega_2 
\nonumber\\&&\sum_{\nu\lambda\mu
  J'M'JM}  (-1)^{\lambda+M+\mu}\,
\left(
\begin{matrix}
J'& J&j \cr M'&-M&0 \cr
\end{matrix} \right)
\left(
\begin{matrix}
J'& J&j \cr \lambda-\mu&\mu-\lambda& 0 \cr
\end{matrix} \right)
 t^{J'M}_{\nu\lambda\mu}(\omega_1,\omega_2)^*\,
t^{JM}_{\nu\lambda\mu}(\omega_1,\omega_2)\,.
\end{eqnarray}

Using the spherical harmonics expansion (\ref{Wij}) should enable one to interpret the
experimental results without resorting to a particular model. This expression
is an analogue to the expansion of the single meson photoproduction cross section in terms
of Legendre polynomials. The coefficients $W_{jm}$
are hermitesch functionals of the partial amplitudes $t^{J M}_{\nu\lambda\mu}$.
They obviously contain the whole information on the
dynamics of the reaction with unpolarized particles and their values may in principle be
extracted from the measurements and compared with model predictions.
The selection rule $W_{jm}=0$ for $j+m=$\,odd may be used for a model independent
partial wave analysis in the low energy region of the reaction, where usually only the
first few waves contribute.

Otherwise an integration over the angles $\theta_\gamma$ and $\phi_\gamma$ gives
the distribution of the events over the Dalitz plot
\begin{equation}\label{Dalitz}
\frac{d^2\sigma}{d\omega_1 d\omega_2}= \pi\,c(W)\sum_{\nu\lambda\mu
JM}\frac{1}{2J+1}|t^{JM}_{\nu\lambda\mu}(\omega_1,\omega_2) |^2\,.
\end{equation}
Thus, as is well known, the partial waves of different $J$ do not interfere in the Dalitz
plot. In spite of its simplicity the expression in Eq.~(\ref{Dalitz}) can hardly be very useful
in reconstructing even the modulae of the amplitudes $t^{JM}_{\nu\lambda\mu}$. Its use
implies that one is able to establish a correspondence between variation of the amplitude
as function of $(\omega_1,\omega_2)$ and a specific value of the total 
angular momentum $J$.
Obviously, for this purpose a detailed model is needed which relates $J$ to 
particular decay channels. In this sense, using the moments $W_{jm}$ should 
be more promising.

It is also clear that the information on the unpolarized differential cross section
only is insufficient for a model independent determination of the amplitudes
$t^{JM}_{\nu\lambda\mu}$. In the general case of photoproduction of two pseudoscalars eight
independent complex functions are required to fix the spin structure of the amplitudes.
Since the overall phase is always arbitrary, one has to measure 15 independent
observables at each kinematical point. However, in certain cases, e.g., when the reaction
is dominated by a single partial wave, using the moments $W_{jm}$
enables one at least to draw a qualitative conclusion with respect to
the partial wave structure. As an illustration, we consider
in the next section the theoretically interesting case of $\pi^0\pi^0$ and $\pi^0\eta$
photoproduction on a proton.

\section{Application to $\gamma p\to \pi^0\pi^0 p$ and $\gamma p\to\pi^0\eta p$}\label{2pi0}

The measured total cross section for $\gamma p\to \pi^0\pi^0 p$
exhibits a rather steep
rise in the energy region below the $D_{13}(1520)$ resonance (see, e.g., \cite{Schum}). At the same
time, the existing models with a dominant contribution from
$D_{13}(1520)$ and a moderate
role of the Roper resonance predict a cross section which increases
rather slowly with increasing energy and
is, therefore, far below the data. It is reasonable to assume that the
almost linear energy
dependence of the data indicates a contribution of  a large fraction of
$s$ waves in the final state. The
main mechanism providing the $s$-wave part in $\pi\pi$ photoproduction is the $\Delta$
Kroll-Ruderman term, appearing after minimal substitution of the electromagnetic interaction
into the $\pi N\Delta$ vertex. This term, however, vanishes in the neutral channel. The
situation is similar to that in single $\pi^0$ photoproduction at low energies. Here the
Kroll-Ruderman does not enter the amplitude, thus leading to a visible suppression of the
cross section for $\gamma p\to\pi^0 p$ in comparison to the $\pi^+$ or $\pi^-$ case.

A possible large contribution of the Roper resonance $P_{11}(1440)$ in the region
$E_\gamma=500-600$~MeV as assumed in Ref.~\cite{Laget} seems to be excluded by more
resent analyses. Furthermore, this assumption should be in disagreement with the
experimental results of Ref.~\cite{Ahren2} for the helicity dependent total cross
section $\Delta\sigma=\sigma_{3/2}-\sigma_{1/2}$. There it was found
that in the energy region
up to at least $E_\gamma=800$~MeV the 3/2 part dominates over the 1/2 part. This means,
that the $P_{11}$ wave, which contributes only to $\sigma_{1/2}$, should be overwhelmed by
the waves with higher spins.

\begin{figure}
\includegraphics[scale=.85]{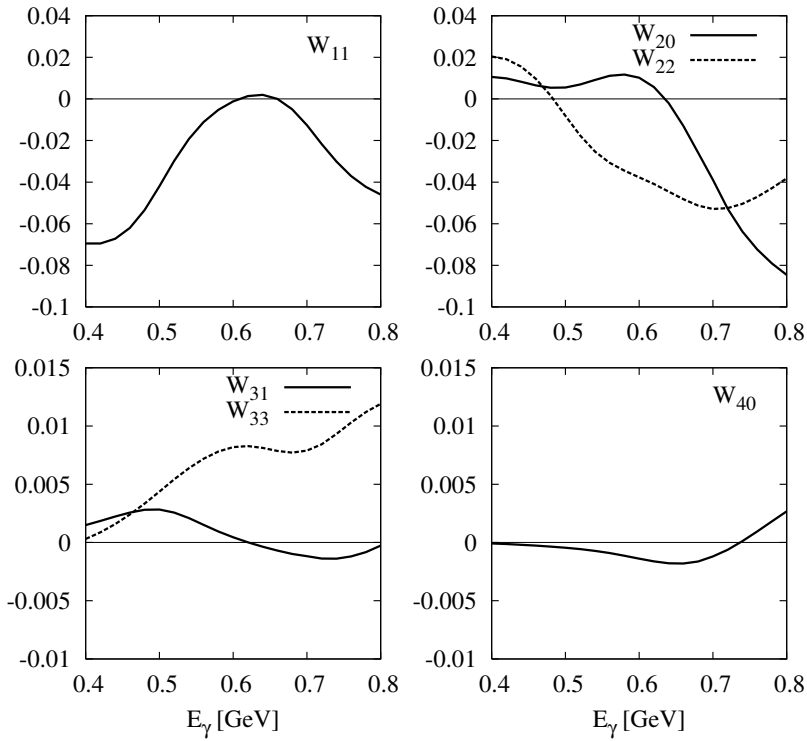}
\caption{The moments $W_{jm}$ for $\gamma p\to\pi^0\pi^0 p$ as functions of the photon
lab energy, normalized such that $W_{00}=1$.} \label{fig2}
\end{figure}

Thus the question concerning the partial wave structure of the amplitude for $\gamma
N\to\pi^0\pi^0 N$ is still open. In order to reveal in this case the mechanism responsible
for an unusually large fraction of the $s$ wave part in the $\pi^0\pi^0$ amplitude, it is useful
to analyse the moments $W_{jm}$ throughout the energy range from
threshold up to the
$D_{13}(1520)$ peak. In order to keep the number of parameters limited, one can use only the
lowest partial waves. Their choice is inspired by the previous isobar model analyses of
Refs.~\cite{Oset,Ochi,FA} showing that only waves with $J\leq 5/2$ are important below
$E_\gamma=1$~GeV.

As an example we show in Fig.~\ref{fig2} the variation of $W_{jm}$ for $j\leq 3$ as
predicted by the $\pi\pi$ model of Ref.~\cite{FA}. The model \cite{FA} is based on a
traditional phenomenological Lagrangean approach with Born and resonance amplitudes
calculated on the tree level. The interaction within the $\pi N$ and $\pi\pi$ pairs is
effectively taken into account via $\Delta$, $\rho$ and $\sigma$. The $\pi\pi
N$ state is then produced through intermediate formation of $\pi\Delta$, $\rho N$ and
$\sigma N$ channels. The contributions from the resonances are parametrized in the usual way
in terms of a Breit-Wigner ansatz with energy-dependent widths. For the parameters of the
model, i.e.\ masses, partial widths and electromagnetic couplings of resonances, the
corresponding average values from the compilation of the Particle Data Group  were used.

In case of $\pi^0\pi^0$ production due to the identity of the two mesons we have an
additional symmetry relation
\begin{equation}\label{identity}
W(\theta_\gamma,\phi_\gamma)=W(\theta_\gamma,2\pi-\phi_\gamma)\,,
\end{equation}
which is a consequence of the symmetry property in Eq.~(\ref{sjm-ident}). The moments for
$j=3,4$ are small as are those for higher values of $j$ which are not shown. In the
region $E=650-800$~MeV the moments $W_{11}$ and $W_{20}$ exhibit a crucial energy
dependence due to the $D_{13}(1520)$ resonance, dominating the reaction $\gamma
p\to\pi^0\pi^0 p$ at this energy. Large values of the moments with $j$ odd indicate the
presence of waves with opposite parities. In particular, the structure in $W_{11}$ is due
to an interference between the wave $J^P=3/2^-$ dominated by $D_{13}(1520)$ and the waves
$J^P=1/2^+$ and $3/2^+$. The latter are saturated, apart from the Roper resonance, by the
Born terms. The contribution of $W_{11}$ becomes minimal in magnitude in the region
around $E_\gamma=650$ MeV, where the real part of the $D_{13}(1520)$ propagator vanishes,
and it interferes weakly with the predominantly real Born amplitudes. Thus, if our notion
about the $\pi^0\pi^0$ photoproduction mechanism is correct we expect a rather small
value of the moments $W_{20}$ and $W_{22}$ and a relatively large value of $W_{11}$ in
the region below the $D_{13}(1520)$ peak.

In this respect we would like to note that according to the fit in Ref.~\cite{Sara}
there must be a large contribution of the resonance $D_{33}(1700)$ to the channel
$\pi^0\pi^0p$ in a wide energy range from the lowest energies up to $E_\gamma=1.4$~GeV. In
particular, inclusion of this resonance into the amplitude explains both the steep rise
of the total cross section below $E_\gamma=700$~MeV and the second peak observed at
$E_\gamma=1.1$~GeV. If the resonance $D_{33}(1700)$ is indeed so important in the
$\pi^0\pi^0$ channel, it should increase the values of $W_{20}$ and $W_{22}$. All in all,
a measurement of these moments will help us to understand the role of $d$-wave resonances
with $J=3/2$ in $\pi^0\pi^0$ photoproduction.

As for $\pi^0\eta$ photoproduction, the partial wave structure of the corresponding 
amplitude was investigated in detail in Refs.~\cite{Horn,Dor,FKLO}. There it was shown
that the $J^P=3/2^-$ wave, containing $D_{33}(1700)$ and probably $D_{33}(1940)$,
apparently dominates the reaction in a wide region from threshold to about $E_\gamma=1.7$
GeV. Other waves, primarily $1/2^+$ and $5/2^+$, manifest themselves in angular
distributions of the final particles mostly via interference with the dominant
$3/2^-$~wave.

\begin{figure}
\includegraphics[scale=.8]{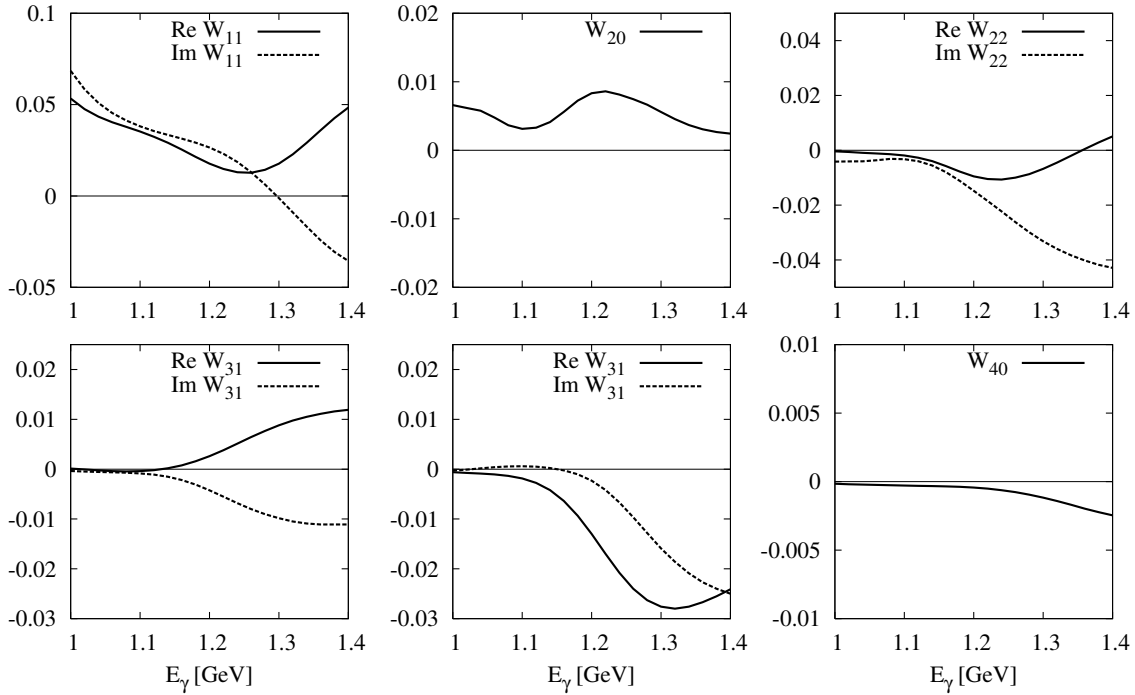}
\caption{Same as in Fig.\,\protect\ref{fig2} for $\gamma p\to\pi^0\eta p$. The dashed
lines represent the imaginary parts.} \label{fig2a}
\end{figure}

In Fig.~\ref{fig2a} we present the energy dependence of the expansion coefficients for
$\gamma p\to\pi^0\eta p$ obtained using the isobar model of Ref.~\cite{FKLO}. Here the
relation (\ref{sjm-ident}) does not hold, so that the moments $W_{jm}$ with $m\ne 0$ have
nonvanishing imaginary parts (dashed lines in Fig.~\ref{fig2a}). The calculation follows
the same line as for the $\pi^0\pi^0$ case. Namely, the final $\pi^0\eta N$ state results
from the two step decay of baryon resonances via the intermediate quasi-two-body channels
$\eta\Delta$ and $\pi^0 S_{11}(1535)$. The parameters of the model were fitted to the angular
distributions of the final particles measured in Ref.\,\cite{Kashev}. The fitting procedure
is described in \cite{FKLO} and the reader is referred to this work for more details.

Firstly, as one can see in Fig.~\ref{fig2a} in spite of the mentioned dominance of the
$3/2^-$ wave, the values of $W_{20}$ and $W_{22}$ are small. This is because of the
closeness of the 3/2 and 1/2 helicity couplings of the resonance $D_{13}(1700)$ (see,
e.g., the discussion in Ref.~\cite{FKLO}). As a result, the hermitian forms of $t^{3/2
M}_{\nu\lambda\mu}$ entering $W_{20}$ and $W_{22}$ according to (\ref{Wjm}) almost cancel
each other. At the same time, we obtain a rather large value of the coefficient $W_{11}$,
mainly determined by the interference between the resonances $D_{33}(1700)$ and
$P_{31}(1750)$. According to these results we may expect that the data for $\pi^0\eta$
will show relatively small values of all moments except for $W_{11}$. If this prediction
is not confirmed by measurements one has to critically review the existing conceptions
about the dynamics of $\pi^0\eta$ photoproduction, based on the results from
Refs.~\cite{Horn,Doring,KashevAphi,Gutz_Is,FKLO}.

\section{Conclusion}\label{conclusion}

Practical methods for the analysis of the partial wave structure of reactions with three
particles in the final state are obviously needed for the study of the
dynamical features of two-meson
photoproduction. The formalism used in the present paper specifies the final $\pi\pi N$
states by means of two c.m.\ energies and two angles,
determining the orientation of
the final state momentum triangle (final state three-particle plane)
with respect to the beam axis. The
partial wave decomposition may then be performed via a transition from the continuum
variables (angles) to the set of discrete variables $JM$ being the total angular
momentum $J$ and its projection $M$ on the normal to the three-particle plane. The
corresponding partial wave amplitudes $t_{\nu\lambda\mu}^{JM}$ contain the whole
information on the production dynamics. We would like to stress the
fact that this method does not involve a
decomposition with respect to the angular momenta of the final
two-body subsystems and is in principle
free from any assumptions about the production mechanism.

In the present paper we have considered only the unpolarized differential cross section.
Although this quantity does not allow a unique determination of the amplitudes
$t_{\nu\lambda\mu}^{JM}$, the information on the angular distribution of the participating
particles can serve to place restrictions on contributions of states with definite
angular momentum and parity. This in turn is crucial for our understanding of the
resonance content of the reaction. For this purpose the
differential cross section has been expanded in terms of spherical
harmonics with coefficients or moments $W_{jm}$ in a manner
similar to the representation of the binary cross section in terms of Legendre
polynomials.

\section*{Acknowledgment}
This work was supported by the Deutsche Forschungsgemeinschaft (SFB 443, SFB 1044) 
and by the Russian Federal programm ``Kadry'' (contract 16.740.11.0469). 
Furthermore, A. Fix thanks the Institut f\"ur Kernphysik in Mainz for their hospitality.

\end{document}